\documentclass[10pt,conference]{IEEEtran}

\usepackage{amsmath,amsthm,amsfonts}
\usepackage{mathtools}
\usepackage{graphicx}
\usepackage{xcolor}
\usepackage{booktabs}
\usepackage{multirow}
\usepackage{makecell}
\usepackage{listings}
\usepackage{enumitem}
\usepackage{pifont}
\usepackage{subcaption}
\usepackage{tikz}
\usetikzlibrary{arrows.meta,positioning,calc}
\usepackage{algorithm}
\usepackage{algpseudocode}
\usepackage{forest}
\usepackage{hyperref} 
\usepackage[capitalise,nameinlink]{cleveref}
\usepackage{newtxmath} 
\usepackage{xspace}
\usepackage{fvextra}
\input{highlighted-code/default-style.tex}

\graphicspath{{../report/}{../report/figures/}{./}{./figures/}}

\newcommand{\chapter}[1]{\section{#1}}

\makeatletter
\newcommand{\abbrev}[1]{\@ifnextchar[{\abbrev@opt{#1}}{#1}}
\newcommand{\abbrev@opt}[2]{#1#2}
\makeatother

\newcommand{\lstsetnumber}[1]{}

\newlist{questions}{enumerate}{2}
\setlist[questions,1]{label=RQ\arabic*.,ref=RQ\arabic*}
\setlist[questions,2]{label=(\alph*),ref=\thequestionsi(\alph*)}

\lstdefinelanguage{diff}{
    basicstyle=\ttfamily\small,
    morecomment=[f][\color{gray}]{@@},
    morecomment=[f][\color{green!50!black}]{+},
    morecomment=[f][\color{red}]{-},
    morecomment=[f][\color{gray}]{---},
    morecomment=[f][\color{gray}]{+++},
}

\newcommand\approach[0]{READU}
\newcommand\readutp[0]{244}
\newcommand\readurepair[0]{217}
\newcommand\readupr[0]{75\%}
\newcommand\readubugsizereport[0]{66}
\newcommand\readubugsizeconfirm[0]{44}
\newcommand\readubugsizefix[0]{26}

\theoremstyle{definition}
\newtheorem{definition}{Definition}[section]

\begin{document}
\title{\approach: Inconsistency-Driven Just-in-Time Detection and Repair of README Bugs}
\author{
    \IEEEauthorblockN{Doehyun Baek}
    \IEEEauthorblockA{CISPA Helmholtz Center for\\
    Information Security\\
    Stuttgart, Germany\\
    doehyunbaek@gmail.com}
    \and
    \IEEEauthorblockN{Kilian Krampf}
    \IEEEauthorblockA{University of Stuttgart\\
    Stuttgart, Germany\\
    kilian@krampf.de}
    \and
    \IEEEauthorblockN{Michael Pradel}
    \IEEEauthorblockA{CISPA Helmholtz Center for\\
    Information Security\\
    Stuttgart, Germany\\
    michael@binaervarianz.de}
}
\maketitle

\begin{abstract}
Repository-level documentation, such as READMEs, is often the first point of contact between users and a repository.
When this documentation is incorrect, users may encounter runtime errors or waste their time debugging.
We call such mistakes in repository-level documentation \emph{README bugs}.
Addressing README bugs is challenging because documentation mixes prose with code, its connection to the source of truth is loose, and finding a bug still leaves developers to craft a repair.
This paper presents \approach{}, an inconsistency-driven technique for just-in-time detection and repair of README bugs.
The key insight behind \approach{} is that README bugs often manifest as inconsistencies between documentation and another source of truth: either repository-internal facts, such as source code, or repository-external facts, such as external dependencies.
\approach{} applies a high-recall commit filter, runs internal and external consistency checkers in parallel, uses an alert judge to remove false positives, and automatically synthesizes documentation patches.
On 6{,}000 recent commits from six popular repositories including Linux and Spring Boot, \approach{} detects \readutp{} true positives with \readupr{} precision, while consuming less than \$0.01 and less than one minute per commit, on average.
Of these true positives, \approach{} correctly repairs \readurepair{}.
We report \readubugsizereport{} found README bugs, of which (so far) \readubugsizeconfirm{} are confirmed and \readubugsizefix{} are fixed.
\end{abstract}

\chapter{Introduction}
\label{s:introduction}

Repository-level documentation---including READMEs, howtos, tutorials, and build or install guides---is often users' first point of contact with a repository.
When it is incorrect, users will likely waste time, e.g., by running into runtime errors triggered by following outdated instructions.
We refer to mistakes in repository-level documentation as \emph{README bugs}, where ``README'' refers not only to files with exactly that name, but to any kind of repository-level documentation.

README bugs are hard to find and fix for three reasons.
First, repository-level documentation mixes prose with code, making correctness difficult to reason about.
Second, repository-level documentation is loosely tied to its sources of truth:
a README may refer to arbitrary code, repository-wide behavior not tied to a specific element, or external facts outside the repository altogether.
Finally, developers can easily miss the needed documentation updates and must still manually craft patches once a bug is found.
Thus, a useful technique should not merely warn, but also suggest mergeable fixes.

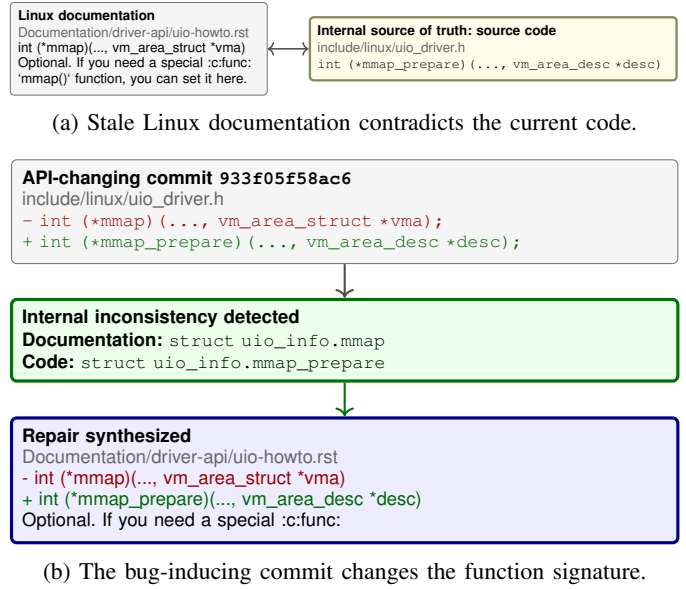
\begin{figure}[t]
    \centering
    \begin{subfigure}[t]{\columnwidth}
        \centering
        \resizebox{\columnwidth}{!}{%
        \begin{tikzpicture}[
            box/.style={draw, rounded corners=2pt, align=left, inner sep=4pt,
                font=\sffamily\scriptsize},
            docs/.style={box, fill=gray!8, draw=black!55, text width=4.4cm},
            code/.style={box, fill=yellow!10, draw=yellow!45!black, very thick,
                text width=6.4cm},
            arrow/.style={->, thick, draw=black!70}
        ]
            \path[draw=none] (0,0) -- (12.2,0);
            \node[docs, anchor=north west] (docs) at (0,0) {\textbf{Linux documentation}\\[0.2ex]
                {\sffamily\color{black!60}Documentation/driver-api/uio-howto.rst}\\[0.2ex]
                {
                \begin{tabular}{@{}l@{}}
                    int (*mmap)(..., vm\_area\_struct *vma)\\
                    Optional. If you need a special :c:func:\\
                    `mmap()` function, you can set it here.
                \end{tabular}}};

            \node[code, anchor=west] (code) at ($(docs.east)+(0.75,0)$) {\textbf{Internal source of truth: source code}\\[0.2ex]
                {\sffamily\color{black!60}include/linux/uio\_driver.h}\\[0.2ex]
                {\ttfamily
                \begin{tabular}{@{}l@{}}
                    int (*mmap\_prepare)(..., vm\_area\_desc *desc)
                \end{tabular}}};

            \draw[<->, thick, draw=black!70] (docs.east) -- (code.west);
        \end{tikzpicture}%
        }
        \caption{Stale Linux documentation contradicts the current code.}
        \label{fig:uio-readu-build}
    \end{subfigure}

    \vspace{0.8em}

    \begin{subfigure}[t]{\columnwidth}
        \centering
        \resizebox{\columnwidth}{!}{%
        \begin{tikzpicture}[
            box/.style={draw, rounded corners=2pt, align=left, inner sep=4pt,
                font=\sffamily\scriptsize},
            source/.style={box, fill=gray!8, draw=black!55, text width=8.6cm},
            readu/.style={box, fill=green!8, draw=green!45!black, very thick,
                text width=8.6cm},
            repair/.style={box, fill=blue!7, draw=blue!55!black, very thick,
                text width=8.6cm},
            arrow/.style={->, thick, draw=black!70}
        ]
            \node[source, anchor=north] (commit) at (0,0) {\textbf{API-changing commit \texttt{933f05f58ac6}}\\[0.2ex]
                {\ttfamily
                \begin{tabular}{@{}l@{}}
                    {\sffamily\color{black!60}include/linux/uio\_driver.h}\\
                    \textcolor{red!60!black}{- int (*mmap)(..., vm\_area\_struct *vma);}\\
                    \textcolor{green!40!black}{+ int (*mmap\_prepare)(..., vm\_area\_desc *desc);}
                \end{tabular}}\\[0.35ex]};
            \node[readu, anchor=north] (detect) at ($(commit.south)+(0,-0.45)$) {\textbf{Internal inconsistency detected}\\[0.2ex]
                \textbf{Documentation:} {\ttfamily struct uio\_info.mmap}\\
                \textbf{Code:} {\ttfamily struct uio\_info.mmap\_prepare}};

            \node[repair, anchor=north] (repair) at ($(detect.south)+(0,-0.45)$) {\textbf{Repair synthesized}\\[0.2ex]
                {
                \begin{tabular}{@{}l@{}}
                    {\sffamily\color{black!60}Documentation/driver-api/uio-howto.rst}\\
                    \textcolor{red!60!black}{- int (*mmap)(..., vm\_area\_struct *vma)}\\
                    \textcolor{green!40!black}{+ int (*mmap\_prepare)(..., vm\_area\_desc *desc)}\\
                    Optional. If you need a special :c:func:
                \end{tabular}}};

            \draw[arrow] (commit.south) -- (detect.north);
            \draw[arrow, draw=green!45!black] (detect.south) -- (repair.north);
        \end{tikzpicture}%
        }
        \caption{The bug-inducing commit changes the function signature.}
        \label{fig:uio-readu-repair}
    \end{subfigure}

    \caption{Stale Linux documentation contradicts the code~\cite{commit_linux_933f05f58}; the internal checker detects it leading to a repair~\cite{commit_linux_de5c46373}.}
    \label{f:motivating-internal}
\end{figure}

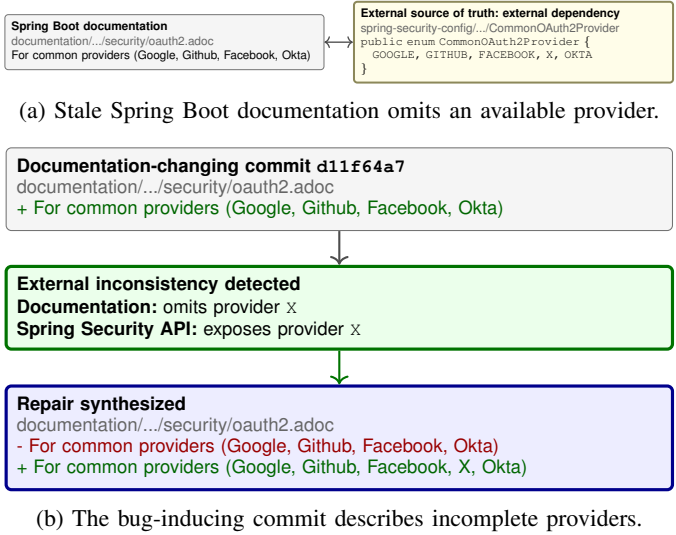
\begin{figure}[!t]
    \centering
    \begin{subfigure}[t]{\columnwidth}
        \centering
        \resizebox{\columnwidth}{!}{%
        \begin{tikzpicture}[
            box/.style={draw, rounded corners=2pt, align=left, inner sep=4pt,
                font=\sffamily\scriptsize},
            doc/.style={box, fill=gray!8, draw=black!55, text width=6.4cm},
            api/.style={box, fill=yellow!10, draw=yellow!45!black, very thick,
                text width=6.4cm},
            arrow/.style={->, thick, draw=black!70}
        ]
            \node[doc, anchor=north west] (docs) at (0,0) {\textbf{Spring Boot documentation}\\[0.2ex]
                {\sffamily\color{black!60}documentation/.../security/oauth2.adoc}\\[0.2ex]
                {
                \begin{tabular}{@{}l@{}}
                    For common providers (Google, Github, Facebook, Okta)
                \end{tabular}}};

            \node[api, anchor=west] (api) at ($(docs.east)+(0.6,0)$) {\textbf{External source of truth: external dependency}\\[0.2ex]
                {\sffamily\color{black!60}spring-security-config/.../CommonOAuth2Provider}\\[0.2ex]
                {\ttfamily
                \begin{tabular}{@{}l@{}}
                    public enum CommonOAuth2Provider \{\\
                    \quad GOOGLE, GITHUB, FACEBOOK, X, OKTA\\
                    \}
                \end{tabular}}};

            \draw[<->, thick, draw=black!70] (docs.east) -- (api.west);
        \end{tikzpicture}%
        }
        \caption{Stale Spring Boot documentation omits an available provider.}
        \label{fig:motivating-external-doc}
    \end{subfigure}

    \vspace{0.8em}

    \begin{subfigure}[t]{\columnwidth}
        \centering
        \resizebox{\columnwidth}{!}{%
        \begin{tikzpicture}[
            box/.style={draw, rounded corners=2pt, align=left, inner sep=4pt,
                font=\sffamily\scriptsize},
            source/.style={box, fill=gray!8, draw=black!55, text width=8.6cm},
            readu/.style={box, fill=green!8, draw=green!45!black, very thick,
                text width=8.6cm},
            repair/.style={box, fill=blue!7, draw=blue!55!black, very thick,
                text width=8.6cm},
            arrow/.style={->, thick, draw=black!70}
        ]
            \node[source, anchor=north] (commit) at (0,0) {\textbf{Documentation-changing commit \texttt{d11f64a7}}\\[0.2ex]
                {
                \begin{tabular}{@{}l@{}}
                    {\sffamily\color{black!60}documentation/.../security/oauth2.adoc}\\
                    \textcolor{green!40!black}{+ For common providers (Google, Github, Facebook, Okta)}\\
                \end{tabular}}};
            \node[readu, anchor=north] (detect) at ($(commit.south)+(0,-0.45)$) {\textbf{External inconsistency detected}\\[0.2ex]
                \textbf{Documentation:} omits provider \texttt{X}\\
                \textbf{Spring Security API:} exposes provider \texttt{X}};

            \node[repair, anchor=north] (repair) at ($(detect.south)+(0,-0.45)$) {\textbf{Repair synthesized}\\[0.2ex]
                {
                \begin{tabular}{@{}l@{}}
                    {\sffamily\color{black!60}documentation/.../security/oauth2.adoc}\\
                    \textcolor{red!60!black}{- For common providers (Google, Github, Facebook, Okta)}\\
                    \textcolor{green!40!black}{+ For common providers (Google, Github, Facebook, X, Okta)}
                \end{tabular}}};

            \draw[arrow] (commit.south) -- (detect.north);
            \draw[arrow, draw=green!45!black] (detect.south) -- (repair.north);
        \end{tikzpicture}%
        }
        \caption{The bug-inducing commit describes incomplete providers.}
        \label{fig:motivating-external-repair}
    \end{subfigure}

    \caption{Stale Spring Boot documentation omits the OAuth2 provider~\cite{ref:spring_boot_oauth2_common_provider_x_doc_stale_fixed}; the external checker detects it leading to a repair~\cite{commit_spring_boot_f7c2c0b2}.}
    \label{fig:motivating-external}
\end{figure}

Figures~\ref{f:motivating-internal} and~\ref{fig:motivating-external} illustrate these challenges.
In Figure~\ref{fig:uio-readu-build}, the Linux documentation mixes a field signature with prose: \texttt{mmap} in \texttt{int (*mmap)(...)} names a \texttt{struct uio\_info} field to implement, whereas \texttt{mmap()} in prose names the memory-mapping operation.
Renaming the field to \texttt{mmap\_prepare} made the signature stale but left the prose reference to \texttt{mmap()} correct.
Figure~\ref{fig:motivating-external-doc} shows a weak link to external dependency: Spring Boot lists common OAuth2 providers but omits \texttt{X}, which Spring Security, an external dependency of Spring Boot, exposes through its \texttt{CommonOAuth2Provider} API.
Both bugs slipped through mature projects with rigorous review standards, motivating automated detection and repair.

Existing approaches do not adequately address these challenges.
Prior work on code-comment inconsistency and API documentation focuses on local code elements~\cite{DBLP:conf/scam/StulovaBGN20, DBLP:conf/acl/PanthaplackelNG20, DBLP:conf/aaai/PanthaplackelLG21, DBLP:conf/icse/RongYLTZSH25, docref, DBLP:conf/icse/ZhouGCHPG17}.
While useful, these techniques are not directly applicable to repository-level documentation that is not tied to a specific code element.
More recent work that considers repository-level documentation, DOCER~\cite{docer} and README-Auto-Update~\cite{readme-auto-update}, still leaves important gaps.
DOCER uses a regular expression to detect a narrow class of stale code element references.
README-Auto-Update uses a fixed LLM pipeline to check whether a README should get modified along with a code change.
However, the technique assumes that there is a single top-level README file (but some projects, such as Linux, have hundreds of repository-level documentation files), does not validate READMEs against project-external facts and lacks support for repairing READMEs.

This paper presents \approach{}, an inconsistency-driven technique for just-in-time detection and repair of README bugs.
The key observation behind \approach{} is that README bugs often manifest as inconsistencies between documentation and another source of truth: either repository-internal facts, such as source code, or repository-external facts, such as external dependencies.
Based on this observation, \approach{} first applies a high-recall commit filter and then runs two LLM agents that check for inconsistencies between a README file and either repository-internal or repository-external facts.
If and only if any of these agents finds two contradictory facts, the approach raises an alert.
After validating potential alerts with an LLM-as-judge mimicking a human developer, the final step of \approach{} is to automatically synthesize a repair for the README bug.

Returning to the motivating examples, \approach{} detects and repairs both bugs by finding the inconsistent facts.
For Linux, the internal checker compares \texttt{Documentation/driver-api/uio-howto.rst} with the changed kernel header, finds that the howto still documents \texttt{struct uio\_info.mmap} while the code exposes \texttt{mmap\_prepare}, and synthesizes the patch shown in Figure~\ref{fig:uio-readu-repair}.
For Spring Boot, the external checker inspects Spring Security's \texttt{CommonOAuth2Provider} API, detects that the documentation omits the supported provider \texttt{X}, and synthesizes the repair shown in Figure~\ref{fig:motivating-external-repair}.
Both patches created by \approach{} have been accepted and merged into Linux's master branch and Spring Boot's main branch, respectively.

We evaluate \approach{} on 6,000 recent commits: 1,000 commits from each of six popular repositories spanning diverse programming languages.
On this dataset, \approach{} detects \readutp{} true positives with \readupr{} precision---3.8$\times$ as many true positives and 12 percentage points higher precision than the strongest baseline, Codex Review---while requiring less than \$0.01 and less than one minute per commit, on average.
Of these true positives, \approach{} correctly repairs \readurepair{}.
We report \readubugsizereport{} found README bugs, of which (so far) \readubugsizeconfirm{} are confirmed and \readubugsizefix{} are fixed.

In summary, this paper makes the following contributions:

\begin{itemize}
    \item \emph{Insight.} We observe that README bugs often manifest as inconsistencies between documentation and repository-internal or repository-external facts.
    \item \emph{Technique.} We present \approach{}, a just-in-time detection and repair technique for README bugs that exploits this observation.
    \item \emph{Evidence.} We empirically show that \approach{} is effective and efficient, and that developers confirm and fix the found bugs.
    \item \emph{Artifacts.} We release our code and data as open source: \url{https://github.com/sola-st/readu}.
\end{itemize}

\section{Approach}
\label{s:approach}

We first define the problem this paper addresses (Section~\ref{s:problem_def}), explain the design rationale of our approach (Section~\ref{s:design_rationale}), give an overview (Section~\ref{s:overview}), and describe each component in detail (Sections~\ref{s:commit_filter}--\ref{s:commit_repair}).

\subsection{Problem Definition}
\label{s:problem_def}

We define a \emph{README bug} as factually incorrect repository-level documentation.
By repository-level documentation, we mean standalone documentation files intended for users or developers at the repository level, rather than comments or docstrings embedded inside source code.
Examples include READMEs, howtos, tutorials, and build or usage guides.
In this paper, repository-level documentation files are identified by paths matching a case-insensitive regex\footnote{\begin{tabular}[t]{@{}l@{}}
\texttt{(\^{}|.*/)((README|INSTALL|BUILD|US(AGE|ING))}\\
\texttt{(\_[A-Z][A-Z])?|DOC(S|UMENTATION)/.*)}\\
\texttt{\textbackslash{}.(MDX?|RST|A(SCII)?DOC|TXT)\$}
\end{tabular}}.

We study the just-in-time setting: the input to the task is a repository transition from a parent commit to a target commit.
The detection output is a set of alerts for README bugs introduced by that transition;
each alert should identify the incorrect documentation statement and the rationale for the decision.
The repair output is a patch against the target commit that removes the README bug.
Detecting bugs at commit time keeps the relevant change small, avoids reporting pre-existing issues, and gives maintainers an opportunity to repair the documentation before users encounter failures.
Automatically repairing bugs saves maintainer effort and provides concrete, actionable suggestions instead of noisy warnings.

\subsection{Design Rationale}
\label{s:design_rationale}

\begin{figure*}[!t]
    \centering
    \includegraphics[width=\textwidth]{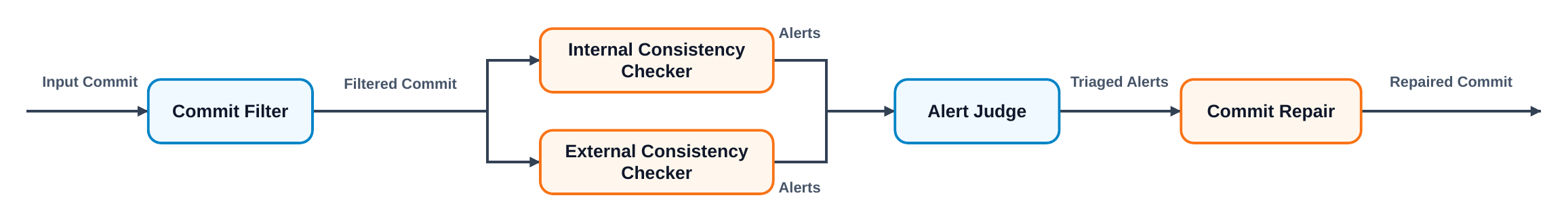}
    \caption{Overview of \approach{}. Blue indicates single-call LLM components; orange indicates LLM-agent components.}
    \label{fig:app-proposed-arch}
\end{figure*}

\approach{} exploits the observation that README bugs often manifest as inconsistencies between documentation and another source of truth.
Hence, we frame detection as consistency checking rather than open-ended bug finding, which offers two advantages.
First, unlike code bugs, README bugs often lack an executable oracle, so detection needs another source of truth.
Second, each inconsistency provides a compact, checkable explanation: a documentation statement and the fact that contradicts it.

We distinguish two kinds of inconsistencies, depending on whether the contradicting fact is repository-internal or repository-external.
We use the following definitions.

\begin{definition}[Internal inconsistency]
\label{def:internal-inconsistency}
An \emph{internal inconsistency} is a contradiction or mismatch between two repository locations where at least one location is repository-level documentation.
\end{definition}

\begin{definition}[External inconsistency]
\label{def:external-inconsistency}
An \emph{external inconsistency} is a repository-level documentation statement or instruction that is out of sync with a concrete fact outside the repository.
\end{definition}

Internal facts include other documentation, code, and configuration in the same repository;
external facts include dependency APIs, tools, and services used by the repository.

\subsection{Overview}
\label{s:overview}

Figure~\ref{fig:app-proposed-arch} gives an overview of \approach.
Given an input commit, the commit filter (Section~\ref{s:commit_filter}) first discards the entire commit if its diff is unlikely to introduce a README bug.
A surviving commit is analyzed in parallel by the internal and external consistency checkers (Sections~\ref{s:internal_checker} and~\ref{s:external_checker}), which compare repository-level documentation against repository-internal and external facts, respectively.
Their alerts are triaged by the alert judge (Section~\ref{s:alert_judge}).
Alerts deemed to be true positives are then sent to the repair component (Section~\ref{s:commit_repair}), which produces a patch that repairs the README bug.

\subsection{Commit Filter}
\label{s:commit_filter}

Popular projects produce many commits, but only a small fraction of them introduce README bugs.
Running the full \approach{} pipeline on every commit would be a substantial waste of computational resources.
Therefore, \approach{} first uses a cheap, high-recall commit filter to discard commits unlikely to introduce README bugs.

The commit filter receives an input commit and first applies a deterministic pass.
If the commit changes a README-like file, the filter keeps the commit without an LLM call, because documentation edits can directly introduce README bugs and should be inspected by the downstream stages.

For the remaining commits, the filter invokes an LLM with the definition of a README bug and the code changes obtained with a Git diff\footnote{\texttt{git show --patch --diff-merges=first-parent}}.
We prompt the LLM as a high-recall filter, as summarized in Figure~\ref{fig:commit-filter-prompt}, because precisely determining whether a commit could introduce a README bug with only Git diff is difficult.

\begin{figure}[t]
    \centering
    \begin{subfigure}[t]{\columnwidth}
        \centering
        \fbox{%
        \begin{minipage}{0.90\columnwidth}
        \footnotesize
        This is a high-recall pre-filter: choose true for plausible stale-README risk ...
        \end{minipage}}
        \caption{Commit filter prompt.}
        \label{fig:commit-filter-prompt}
    \end{subfigure}

    \vspace{0.4em}

    \begin{subfigure}[t]{\columnwidth}
        \centering
        \fbox{%
        \begin{minipage}{0.90\columnwidth}
        \footnotesize
        A true positive is an alert that identifies a real README/documentation bug introduced by the commit: ... 

        A false positive is speculative, not introduced by the commit, not tied to documentation, not supported by the supplied evidence, ...
        \end{minipage}}
        \caption{Alert judge prompt.}
        \label{fig:alert-judge-rubric}
    \end{subfigure}

    \vspace{0.4em}

    \begin{subfigure}[t]{\columnwidth}
        \centering
        \fbox{%
        \begin{minipage}{0.90\columnwidth}
        \footnotesize
        Prefer editing README-like documentation files...
        
        Keep changes minimal and directly tied to the supplied alerts.
        \end{minipage}}
        \caption{Commit repair prompt.}
        \label{fig:commit-repair-prompt}
    \end{subfigure}
    \caption{Prompt summaries for the commit filter, alert judge, and commit repair components. Full prompts for these and other components are in the artifact.}
    \label{fig:component-prompt-summaries}
\end{figure}

\begin{figure}[t]
    \centering
    \begin{subfigure}[t]{\columnwidth}
        \centering
        \input{highlighted-code/commit-reject.tex}
        \caption{Should reject: Javadoc typo in Spring Boot~\cite{commit_spring_boot_29b82310}.}
        \label{fig:commit-filter-example-negative}
    \end{subfigure}

    \vspace{0.5em}

    \begin{subfigure}[t]{\columnwidth}
        \centering
        \input{highlighted-code/commit-accept.tex}
        \caption{Should accept: stale README in React~\cite{commit_react_848e0e3}.}
        \label{fig:commit-filter-example-positive}
    \end{subfigure}
    \caption{Commit-filter examples.}
    \label{fig:commit-filter-example}
\end{figure}

Figure~\ref{fig:commit-filter-example} shows two commit-filter examples.
The first should be rejected because it only fixes a typo in source-level Javadoc and does not change behavior, commands, paths, or configuration.
The second should be kept: the commit replaces the \texttt{extends}-based setup with the configuration style introduced in ESLint v9.0.0\footnote{\url{https://eslint.org/blog/2024/04/eslint-v9.0.0-released/}}.
Indeed, as determined by a later step of \approach{}, the existing README file becomes stale by this commit, as it still documents an \texttt{extends}-based setup.

\subsection{Internal Consistency Checker}
\label{s:internal_checker}

Many README bugs are caused by repository-internal changes: code, configuration, paths, commands, or other documentation evolve while the corresponding documentation stays unchanged.
These bugs require inspecting the post-commit repository, not just the changed diff.
However, placing the entire repository into a single LLM context is neither effective nor efficient: context bloat can make the model miss the relevant contradiction and would waste substantial LLM resources.

To explore repository state and possible inconsistencies flexibly, we implement the internal consistency checker as an LLM agent.
Starting from the post-commit checkout, the agent first inspects changed files and the diff with \texttt{git}, then performs targeted searches over documentation and repository symbols.
It detects inconsistencies between repository-level documentation and repository-internal facts.
The agent reads only the relevant documentation and code snippets, compares them, and returns JSON alerts in which each alert identifies the documentation location and the contradicting repository location.

For the Linux bug in Figure~\ref{f:motivating-internal}, the checker run proceeds as follows.
The agent begins with the changed files and diff, observes that the commit changes UIO code and \texttt{include/linux/uio\_driver.h}, and uses these terms to search repository documentation.
It opens \texttt{Documentation/driver-api/uio-howto.rst}, narrows to the bullet documenting the \texttt{struct uio\_info} callback, and compares the documented callback name and signature with the updated header.
It then returns a JSON alert that pairs the stale documentation lines with the contradicting header lines.

\subsection{External Consistency Checker}
\label{s:external_checker}

While the internal consistency checker can validate README bugs against repository-internal facts, it cannot validate README bugs that involve repository-external facts.
However, some README bugs are caused by changes to external facts, such as APIs of third-party packages, or tools and services the analyzed repository uses.
These facts may change outside the repository or differ from what the repository examples assume.

The external consistency checker therefore detects inconsistencies between repository-level documentation and external facts.
We implement it as an LLM agent whose prompt instructs it to verify such facts with external checks, including \texttt{curl} for URLs and install commands, package-manager queries such as \texttt{npm view}, and public release or API queries.
The agent may also download and inspect published artifacts, such as npm packages, when metadata alone is insufficient.
These tools are necessary because the relevant source of truth may be outside the repository and may change independently of the commit.
We prompt the agent to return alerts that identify the documentation location, the contradicting external fact, and, when available, an authoritative URL for independent verification.

For the Spring Boot bug in Figure~\ref{fig:motivating-external}, the checker run proceeds as follows.
The agent follows the changed OAuth2 documentation to \texttt{OAuth2ClientPropertiesMapper}, which delegates common providers to Spring Security, which is an external dependency of Spring Boot.
It checks the declared Spring Security version, inspects Spring Security's \texttt{CommonOAuth2Provider} implementation for that version, verifies that \texttt{X} is supported, and returns an alert for the incomplete provider lists.

\begin{figure}[t]
    \centering
    \begin{subfigure}[t]{\columnwidth}
        \centering
        \begin{tikzpicture}[
            font=\footnotesize,
            node distance=0.22cm,
            alert/.style={draw, rounded corners, fill=gray!6, align=left, text width=0.88\linewidth, inner sep=4pt},
            decision/.style={draw, rounded corners, align=center, text width=0.72\linewidth, inner sep=4pt, font=\footnotesize\bfseries},
            reject/.style={decision, fill=red!6, draw=red!60!black},
            reason/.style={draw, rounded corners, fill=blue!4, align=left, text width=0.88\linewidth, inner sep=4pt},
            arrow/.style={-{Stealth[length=1.8mm]}, thick, draw=gray!70}
        ]
            \node[alert] (ag-alert) {\textbf{Input alert.} External checker: frontend README allows Node.js 16.10+, but \texttt{package.json} requires Node.js 22.x.};
            \node[reject, below=of ag-alert] (ag-decision) {Judge: reject};
            \node[reason, below=of ag-decision] (ag-reason) {\textbf{Reason.} The mismatch existed before the analyzed commit, so it was not introduced by the commit.};
            \draw[arrow] (ag-alert) -- (ag-decision);
            \draw[arrow] (ag-decision) -- (ag-reason);
        \end{tikzpicture}
        \caption{Rejected alert example from AutoGPT: pre-existing issue~\cite{commit_autogpt_36fb1ea}.}
        \label{fig:alert-judge-example-pre-existing}
    \end{subfigure}

    \vspace{0.5em}

    \begin{subfigure}[t]{\columnwidth}
        \centering
        \begin{tikzpicture}[
            font=\footnotesize,
            node distance=0.22cm,
            mini-alert/.style={draw, rounded corners, fill=gray!6, align=center, text width=0.25\linewidth, minimum height=1.12cm, inner sep=4pt, font=\scriptsize},
            mini-decision/.style={draw, rounded corners, align=center, text width=0.25\linewidth, minimum height=0.50cm, inner sep=4pt, font=\scriptsize\bfseries},
            keep/.style={mini-decision, fill=green!7, draw=green!40!black},
            duplicate/.style={mini-decision, fill=orange!10, draw=orange!70!black},
            reason/.style={draw, rounded corners, fill=blue!4, align=left, text width=0.88\linewidth, inner sep=4pt},
            arrow/.style={-{Stealth[length=1.8mm]}, thick, draw=gray!70}
        ]
            \node[mini-alert, xshift=-0.18\linewidth] (tf-internal) {\textbf{Input alert 1}\\Internal\\old XLA\\\texttt{BUILD} path};
            \node[mini-alert, xshift=0.18\linewidth] (tf-external) {\textbf{Input alert 2}\\External\\same link\\returns 404\vphantom{\texttt{BUILD}}};
            \node[keep, below=of tf-internal] (tf-keep) {Judge: keep};
            \node[duplicate, below=of tf-external] (tf-duplicate) {Judge: duplicate};
            \coordinate[below=0.35cm of $(tf-keep.south)!0.5!(tf-duplicate.south)$] (tf-reason-anchor);
            \node[reason, anchor=north] (tf-reason) at (tf-reason-anchor) {\textbf{Reason.} Both alerts describe the same stale link; the internal alert gives clearer repository evidence for repair.};
            \draw[arrow] (tf-internal) -- (tf-keep);
            \draw[arrow] (tf-external) -- (tf-duplicate);
            \draw[arrow] (tf-keep.south) -- (tf-reason.north);
            \draw[arrow] (tf-duplicate.south) -- (tf-reason.north);
        \end{tikzpicture}
        \caption{Duplicate alerts example from TensorFlow~\cite{commit_tensorflow_0d5d53b}.}
        \label{fig:alert-judge-example-duplicate}
    \end{subfigure}

    \caption{Examples of alert judge decisions.}
    \label{fig:alert-judge-example}
\end{figure}
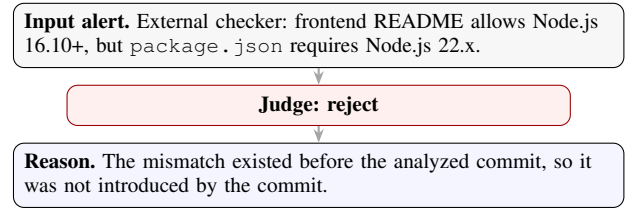
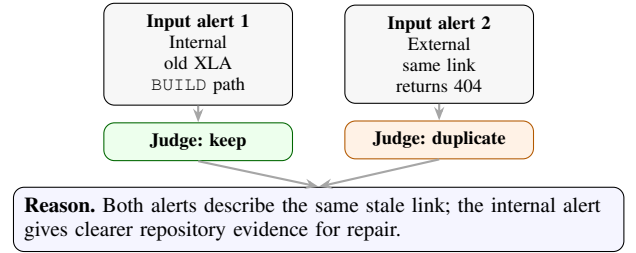

\subsection{Alert Judge}
\label{s:alert_judge}

The internal and external checkers are intentionally high-recall and use different sources of evidence.
Forcing each checker to stay within its nominal scope reduced recall, because real README bugs can straddle repository and external evidence.
\approach{} therefore lets both checkers report plausible alerts and delegates final triage to an alert judge.

We implement the judge as a single LLM call that takes as input the candidate alerts, the post-commit repository-level documentation, and the commit diff.
For each alert, the judge returns a decision and rationale that determine whether the repair component receives the alert.
Figure~\ref{fig:alert-judge-rubric} summarizes the rubric specified in the judge prompt.

First, the alert judge filters false positives.
It keeps only alerts that correspond to actionable README bugs introduced by the input commit and rejects alerts that are speculative, unsupported by the supplied evidence, not tied to repository-level documentation, or describe issues that already existed before the input commit.

Second, the alert judge deduplicates alerts.
Because the internal and external checkers can report the same underlying README bug from different evidence paths, the judge groups duplicate alerts, keeps the clearer alert, and marks the others as duplicates.
Thus, the repair component receives at most one request per bug.

Figure~\ref{fig:alert-judge-example} shows examples of both decisions.
For AutoGPT, the external checker reports that the frontend README allows Node.js 16.10+ while \texttt{package.json} requires Node.js 22.x; the judge rejects it because this mismatch existed before the input commit.
For TensorFlow, the internal and external checkers both report the same stale XLA \texttt{BUILD} link, so the judge keeps the internal alert and marks the external alert as a duplicate.

\subsection{Commit Repair}
\label{s:commit_repair}

Detection is most useful when maintainers can act on it immediately, but producing a safe patch still requires navigating the repository and preserving documentation style.
Recent work on LLM-based software-engineering agents shows that agents can navigate repositories and synthesize patches for real-world software issues~\cite{repairagent,sweagent,autocoderover,googleapr}.
Compared with general software repair, this task is easier because accepted alerts already identify the stale documentation, the contradicting fact, and a concise rationale.
Still, the required patch can still be multi-hunk when the same stale fact appears in several places.

Motivated by this progress, the commit repair component employs an LLM agent to produce a patch for each true-positive alert accepted by the alert judge.
Figure~\ref{fig:commit-repair-prompt} summarizes the repair policy specified in the prompt.
The agent starts from the accepted alert, inspects only the relevant documentation and supporting evidence, and produces a minimal documentation-oriented patch to remove the inconsistency.
We prompt the agent to prefer README-like documentation files, keep changes minimal and directly tied to the supplied alerts, preserve surrounding style and terminology, and avoid unrelated cleanups.
Before finishing, it rereads the changed snippets, runs \texttt{git diff} on the modified files, and returns a JSON object containing a summary, modified files, and the exact patch.
It must preserve unrelated content and formatting.

For the motivating Linux bug (Figure~\ref{f:motivating-internal}), the synthesized repair updates the UIO howto to use \texttt{mmap\_prepare} instead of the stale \texttt{mmap} callback.
For the motivating Spring Boot bug (Figure~\ref{fig:motivating-external}), the synthesized repair adds \texttt{X} to the incomplete provider lists.

\chapter{Evaluation}
\label{s:evaluation}

We evaluate \approach{} with the following research questions:

\begin{questions}
    \item \textbf{Effectiveness:} How effective is \approach{} at detecting and repairing README bugs?\label{rq:effective}
    \item \textbf{Feedback from real-world developers:} How do maintainers respond to README bugs found by \approach{}?\label{rq:inthewild}
    \item \textbf{Efficiency:} How efficient is \approach{} in terms of time, token consumption, and monetary cost?\label{rq:efficient}
    \item \textbf{Component usefulness:} How much do individual components of \approach{} contribute to the overall results?\label{rq:ablation}
\end{questions}

\subsection{Experimental Setup}
\label{s:evaluation-setup}

\subsubsection{Dataset}

\paragraph{Repository Selection}

We select GitHub repositories from six language groups: C, C++, Java, Go, Python, JavaScript/TypeScript.
Repositories must have at least $500$ stars, must not be archived, must have been created before 2025-10-06, must have been updated since 2026-01-01, must contain at least one repository-level documentation, and must have that language group account for at least $50\%$ of the repository.
We query the GitHub API for candidates ordered descendingly by stars.
After manually excluding $10$ unsuitable repositories ($6$ educational resources, $3$ awesome lists, and $1$ with a non-English README), we retain the highest-ranked remaining repository per language group.

\paragraph{Commit Selection}
For each selected repository, we select commits on its default branch from 2025-04-07 to 2026-04-07.
This results in $113{,}393$ commits total.
Because some repositories contain many selected commits (e.g., Linux with $88{,}707$ commits), we evaluate the $1{,}000$ most recent commits per repository.
Each approach receives one repository-commit pair per task.
The approach analyzes the transition from the parent commit to the target commit and raises alerts for README bugs introduced by that transition.
For merge commits, we use the first parent as the parent commit.
Table~\ref{tab:evaluation-dataset} summarizes the repositories, commits, and evaluated commits under consideration.

\begin{table}[t]
    \caption{Evaluation dataset.}
    \label{tab:evaluation-dataset}
    \centering
    \footnotesize
    \resizebox{\columnwidth}{!}{
        \begin{tabular}{@{}llrr@{}}
            \toprule
            \textbf{Repository} & \textbf{Language} & \textbf{Commits} & \textbf{Evaluated Commits} \\
            \midrule
            Ollama~\cite{ollama_ollama} & Go & $1{,}140$ & $1{,}000$ \\
            React~\cite{react_react} & JavaScript/TypeScript & $1{,}223$ & $1{,}000$ \\
            AutoGPT~\cite{autogpt} & Python & $1{,}642$ & $1{,}000$ \\
            Spring Boot~\cite{spring_boot} & Java & $6{,}790$ & $1{,}000$ \\
            TensorFlow~\cite{tensorflow} & C++ & $13{,}891$ & $1{,}000$ \\
            Linux~\cite{linux} & C & $88{,}707$ & $1{,}000$ \\
            \midrule
            \textbf{Total} & & $113{,}393$ & $6{,}000$ \\
            \bottomrule
        \end{tabular}
    }
\end{table}

\subsubsection{Metrics}

We derive effectiveness metrics from manual annotations.
For detection, annotators label each alert as a true positive if it reports a README bug introduced by the evaluated commit, and as a false positive otherwise.
We report true positives, false positives, and precision for each approach.
We do not report recall because we do not construct a complete ground truth of all README bugs introduced by the evaluated commits.
For repair, annotators label each generated patch as correct, incorrect, or unsure.
We use unsure only for repair because patch correctness can depend on project-specific behavior or complex commands that are difficult to validate reliably; we conservatively count unsure as incorrect for both agreement and reported metrics.
For efficiency, we report wall-clock time, token consumption, and monetary cost.

Two authors independently annotated overlapping samples to assess annotation reliability with annotation rubrics.
We report Gwet's $AC1$~\cite{gwet2008computing} because it is robust to skewed label distributions.
After an initial calibration phase ($219$ detection and $60$ repair labels; $AC1=0.692$ and $0.667$), the authors resolved disagreements and refined the annotation rubrics.
In the final pass, agreement improved to $AC1=0.853$ for detection ($270/292$ agreements, $92.47\%$) and $AC1=0.848$ for repair ($132/150$ agreements, $88.00\%$).
Using the finite-population sample-size formula ($N_{\text{det}}=849$, $N_{\text{rep}}=244$, $p=0.5$)~\cite{krejcie_determining_1970}, these overlaps are sufficient for a $\pm5\%$ margin of error at $95\%$ confidence.
For the reported metrics, we use the annotations from the first author.

\subsubsection{Baselines}
We compare \approach{} against five detection baselines: three general-purpose techniques and two techniques specialized for README bugs.
\begin{itemize}[leftmargin=*,nosep]
    \item \textbf{Single LLM invocation} tests whether one structured LLM call is sufficient.
    \item \textbf{Mini-SWE-agent} represents a general-purpose agentic software-engineering baseline~\cite{sweagent,noauthor_swe-agentmini-swe-agent_2026}.
    \item \textbf{Codex Review} represents a general-purpose code change review assistant~\cite{openai_codex_review}.
    \item \textbf{DOCER} is a regex-based detector for outdated code element references in repository-level documentation~\cite{docer}.
    \item \textbf{README-Auto-Update} is an LLM-based technique that, given a pull request, predicts whether a README update is needed and localizes the README sections that likely require updates~\cite{readme-auto-update}.
\end{itemize}

For detection, we run Codex Review, DOCER, and README-Auto-Update as-is; for Single LLM invocation and Mini-SWE-agent, we prompt them with our README-bug definition.
For repair, we only evaluate \approach{} because the baselines do not synthesize documentation patches.

\subsubsection{Implementation}

We implement \approach{} and the evaluation harness in Python~3.13.13.
The agent-based components---the internal consistency checker, external consistency checker, and commit repair---are implemented with pi-coding-agent~v0.78.0~\cite{pi_coding_agent}.
The commit filter and alert judge are implemented as structured single-call LLM invocations.
For the experiments in this paper, all LLM invocations across the baselines and \approach{} use the same open-weight DeepSeek V4 Flash model with high reasoning effort~\cite{deepseekai2026deepseekv4}.

\subsection{RQ1: Effectiveness}
\label{s:effectiveness}

\subsubsection{Quantitative Results}

\begin{table}[!t]
    \caption{Detection evaluation results.}
    \label{tab:big-evaluation}
    \centering
    \scriptsize
    \setlength{\tabcolsep}{2pt}
    \renewcommand{\arraystretch}{1.05}
    \resizebox{\columnwidth}{!}{
        \begin{tabular}{@{}lrrrrrr@{}}
            \toprule
            \multirow{2}{*}{\textbf{Technique / variant}} & \multicolumn{3}{c}{\textbf{Effectiveness}} & \multicolumn{3}{c}{\textbf{Avg. per commit}} \\
            \cmidrule(lr){2-4}\cmidrule(l){5-7}
            & \textbf{TP} & \textbf{FP} & \textbf{Precision (\%)} & \textbf{Time} & \textbf{Tokens} & \textbf{Cost} \\
            \midrule
            \multicolumn{7}{@{}l}{\textit{General-purpose baselines}} \\
            Single LLM invocation & 29 & 18 & 62\% & 122.18s & 397k & \$0.0102 \\
            Mini-SWE-agent & 63 & 45 & 58\% & 24.18s & 22k & \$0.0010 \\
            Codex Review & 64 & 38 & 63\% & 130.93s & 504k & \$0.0073 \\
            \midrule
            \multicolumn{7}{@{}l}{\textit{Task-specific baselines}} \\
            DOCER & 0 & 3 & 0\% & 86.52s & N/A & N/A \\
            README-Auto-Update & 7 & 29 & 19\% & 10.66s & 5k & \$0.0002 \\
            \midrule
            \multicolumn{7}{@{}l}{\textit{Approach}} \\
            \approach{} & 244 & 81 & 75\% & 47.08s & 337k & \$0.0066 \\
            \midrule
            \multicolumn{7}{@{}l}{\textit{Ablations}} \\
            w/o internal checker & 74 & 35 & 68\% & 34.68s & 154k & \$0.0033 \\
            w/o external checker & 204 & 51 & 80\% & 40.55s & 192k & \$0.0044 \\
            w/o alert judge & 275 & 180 & 60\% & 45.30s & 318k & \$0.0044 \\
            \bottomrule
        \end{tabular}
    }
\end{table}

\begin{table}[!t]
    \caption{True-positive detection counts by repository, with per-repository false positives and precision for \approach{}.}
    \label{tab:per-repo-effectiveness}
    \centering
    \scriptsize
    \setlength{\tabcolsep}{2pt}
    \renewcommand{\arraystretch}{1.05}
    \resizebox{\columnwidth}{!}{
        \begin{tabular}{@{}lrrrrrr@{}}
            \toprule
            \textbf{Technique / metric} & \textbf{Ollama} & \textbf{AutoGPT} & \textbf{Linux} & \textbf{Spring Boot} & \textbf{React} & \textbf{TensorFlow} \\
            \midrule
            Single LLM invocation & 16 & 7 & 0 & 3 & 3 & 0 \\
            Mini-SWE-agent & 27 & 21 & 4 & 4 & 7 & 0 \\
            Codex Review & 39 & 3 & 5 & 16 & 1 & 0 \\
            DOCER & 0 & 0 & 0 & 0 & 0 & 0 \\
            README-Auto-Update & 7 & 0 & 0 & 0 & 0 & 0 \\
            \midrule
            \multicolumn{7}{@{}l}{\textit{\approach{}}} \\
            \quad TP & 120 & 75 & 26 & 17 & 3 & 3 \\
            \quad FP & 39 & 17 & 6 & 7 & 11 & 1 \\
            \quad Precision & 75\% & 82\% & 81\% & 71\% & 21\% & 75\% \\
            \bottomrule
        \end{tabular}
    }
\end{table}

\begin{table}[t]
    \caption{Repair evaluation results.}
    \label{tab:repair-performance}
    \centering
    \footnotesize
    \resizebox{\columnwidth}{!}{
        \begin{tabular}{@{}lcccc@{}}
            \toprule
            \multirow{2}{*}{\textbf{Technique}} & \multirow{2}{*}{\textbf{Correct repairs / Targets}} & \multicolumn{3}{c}{\textbf{Avg. per target}} \\
            \cmidrule(l){3-5}
            & & \textbf{Time (s)} & \textbf{Tokens} & \textbf{Cost} \\
            \midrule
            \approach{} & 217/244 (89\%) & 89.7 & 107.7k & \$0.0081 \\
            \bottomrule
        \end{tabular}
    }
\end{table}

The effectiveness columns of Table~\ref{tab:big-evaluation} show that \approach{} detects $\readutp{}$ README bugs with \readupr{} precision.
Table~\ref{tab:per-repo-effectiveness} shows that these results are not driven by a single repository: apart from React, per-repository precision remains between $71\%$ and $82\%$.
Most true positives come from Ollama and AutoGPT ($196/\readutp{}$), but \approach{} also finds README bugs in Linux, Spring Boot, React, and TensorFlow.

Table~\ref{tab:repair-performance} shows that the repair component produces correct patches for $217$ of $\readutp{}$ targets ($89\%$).
Combining detection and repair is useful because accepted alerts already provide the repair component with the affected documentation location and the contradicting fact.
The motivating repairs in Figures~\ref{f:motivating-internal} and~\ref{fig:motivating-external} are representative:
the Linux patch updates the stale \texttt{mmap} field reference, while the Spring Boot patch adds the missing \texttt{X} provider.
Patch ingredients for both patches are already contained in the alerts.

\subsubsection{Characterization of True Positives}

We characterize \approach{}'s true positives by theme.
The $\readutp{}$ true positives comprise five README-bug themes and one typo; Figure~\ref{fig:readu-unique-examples} shows one representative bug per theme.
\begin{itemize}[leftmargin=*,nosep]
    \item \textbf{Implementation--documentation drift} ($81$): documentation states values or behavior that no longer match the implementation, rendering the documentation internally inconsistent.
    The example shows an image-generation README still claiming a fixed $30$-step setting, while the CLI implementation defines the default as $9$ steps.
    \item \textbf{Incorrect external references} ($63$): documentation records externally verifiable references or requirements---such as live URLs, package names, model registry entries, or supported platforms---that do not match the external source of truth.
    The example pipes an install script from a URL that returns HTTP 404.
    \item \textbf{Missing API/feature documentation} ($48$): newly added fields, parameters, endpoints, permissions, or features are omitted. The example lists \texttt{tool\_calls} and \texttt{tool\_name} but omits \texttt{tool\_call\_id}.
    \item \textbf{Invalid usage instructions} ($27$): examples, commands, or setup steps are syntactically invalid, refer to missing files, or use the wrong command.
    The example omits the required \texttt{-tags mlx} build flag for source guarded by \texttt{//go:build mlx}.
    \item \textbf{Cross-document, translation, or naming drift} ($24$): related documents, translations, links, anchors, or names diverge after a partial update. The example leaves the Italian Linux requirements at Rust $1.78.0$ after the English page moves to Rust $1.85.0$.
\end{itemize}

\begin{figure}[t]
    \centering
    \captionsetup[subfigure]{skip=2pt}
    \begin{subfigure}[t]{\columnwidth}
        \begin{minipage}[t]{0.495\columnwidth}
            \input{highlighted-code/readu-fixed-steps.tex}
        \end{minipage}
        \hfill
        \begin{minipage}[t]{0.495\columnwidth}
            \input{highlighted-code/readu-steps-code.tex}
        \end{minipage}
        \caption{Implementation--documentation drift~\cite{commit_ollama_25849400}}
    \end{subfigure}
    \par\vspace{0.15em}
    \begin{subfigure}[t]{\columnwidth}
        \begin{minipage}[t]{0.495\columnwidth}
            \input{highlighted-code/readu-install-doc.tex}
        \end{minipage}
        \hfill
        \begin{minipage}[t]{0.495\columnwidth}
            \input{highlighted-code/readu-install-check.tex}
        \end{minipage}
        \caption{Incorrect external references~\cite{commit_ollama_5267d31d}}
    \end{subfigure}
    \par\vspace{0.15em}
    \begin{subfigure}[t]{\columnwidth}
        \begin{minipage}[t]{0.495\columnwidth}
            \input{highlighted-code/readu-api-doc.tex}
        \end{minipage}
        \hfill
        \begin{minipage}[t]{0.495\columnwidth}
            \input{highlighted-code/readu-api-code.tex}
        \end{minipage}
        \caption{Missing API/feature documentation~\cite{commit_ollama_809b9c68}}
    \end{subfigure}
    \par\vspace{0.15em}
    \begin{subfigure}[t]{\columnwidth}
        \begin{minipage}[t]{0.495\columnwidth}
            \input{highlighted-code/readu-build-doc.tex}
        \end{minipage}
        \hfill
        \begin{minipage}[t]{0.495\columnwidth}
            \input{highlighted-code/readu-build-code.tex}
        \end{minipage}
        \caption{Invalid usage instructions~\cite{commit_ollama_33ee7168}}
    \end{subfigure}
    \par\vspace{0.15em}
    \begin{subfigure}[t]{\columnwidth}
        \begin{minipage}[t]{0.495\columnwidth}
            \input{highlighted-code/readu-translation-it.tex}
        \end{minipage}
        \hfill
        \begin{minipage}[t]{0.495\columnwidth}
            \input{highlighted-code/readu-translation-en.tex}
        \end{minipage}
        \caption{Cross-document or translation drift~\cite{commit_linux_b28711ac}}
    \end{subfigure}
    \caption{Representative \approach{} true positives by theme. Each subfigure contrasts the documented claim (left) with the contradicting internal or external fact (right).}
    \label{fig:readu-unique-examples}
\end{figure}

\subsubsection{Characterization of False Positives}

Although \approach{} achieves \readupr{} precision, its false positives reveal the remaining challenges in distinguishing newly introduced README bugs from plausible but non-actionable inconsistencies.
We categorize its $81$ false positives into five causes:
\begin{itemize}[leftmargin=*,nosep]
    \item \textbf{Pre-existing issues} ($32$): the alert identifies a plausible inconsistency, but it was not introduced by the evaluated commit.
    While such alerts may still be useful to developers, we count them as false positives because our approach aims to detect bugs introduced by the evaluated commit.
    \item \textbf{Misinterpreted implementation semantics or context} ($18$): the checker misreads code semantics (e.g., inheritance or global behavior) or the intended scope of the code change.
    \item \textbf{Intentionally non-exhaustive documentation} ($14$): the documentation is an example, partial list, or high-level guide rather than a complete reference.
    \item \textbf{Benign equivalence, wording, or optionality} ($12$): the checker treats equivalent syntax, wording differences, optional steps, or harmless duplication as contradictions.
    \item \textbf{Incorrect external evidence or source-of-truth selection} ($5$): the external checker relies on incomplete evidence, stale availability checks, or the wrong source of truth.
\end{itemize}

\subsubsection{Characterization of Unsuccessful Repairs}

For the 27 repair targets not marked correct, manual annotation marked 11 patches as unsure because they involved complex commands that were difficult to validate reliably, and 16 patches as incorrect.
The 16 manually annotated incorrect repairs fall into four categories:
\begin{itemize}[leftmargin=*,nosep]
    \item \textbf{Incomplete or under-specified repair} ($5$): the patch moves in the right direction but omits important qualifiers, links, command details, or device mappings.
    \item \textbf{Wrong replacement or root-cause mismatch} ($5$): the patch edits the wrong value, adds redundant guidance, or fixes a surface symptom.
    \item \textbf{No usable repair generated} ($4$): the repair run produced no usable patch because of LLM failures.
    \item \textbf{Overly destructive repair} ($2$): the patch removes or weakens useful documentation instead of preserving the intention.
\end{itemize}

\begin{figure}[t]
    \centering
    \includegraphics[width=\columnwidth]{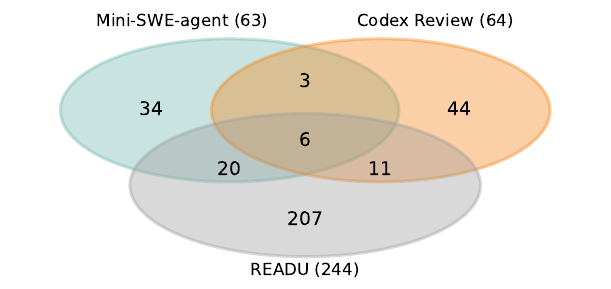}
    \caption{Venn diagram of true positives across agentic approaches.}
    \label{fig:detection-venn}
\end{figure}

\subsubsection{Comparison with Baselines}

\approach{} detects substantially more README bugs than all baselines while maintaining the highest precision among complete techniques (Table~\ref{tab:big-evaluation}).
The task-specific baselines cover narrower problem classes.
DOCER's regex-based matching is too brittle: its only alerts were two false positives, including one that matched a deleted Flux $20\,\mathrm{GB}$ VRAM estimate to an unrelated README model-size entry\footnote{\url{https://github.com/ollama/ollama/commit/3b3bf6}}.
README-Auto-Update is a fixed LLM pipeline for a known target README rather than a full-fledged LLM agent.
Targeting only the root README misses most bugs elsewhere: $228/244$ of \approach{}'s true positives are in non-root repository-level documentation.
Even when restricted to root READMEs, \approach{} finds $16$ true positives ($14$ in Ollama and $2$ in AutoGPT), compared with README-Auto-Update's seven, all in Ollama.
Single LLM invocation improves upon task-specific baselines with 29 true positives with $62\%$ precision, but still lags substantially behind three agentic approaches evaluated.

Figure~\ref{fig:detection-venn} shows substantial non-overlap among the agentic approaches, suggesting that they find different classes of README bugs.
Mini-SWE-agent is effective at finding commands that fail when run, and Codex Review is strong on small localized documentation-quality nits.
For Mini-SWE-agent, invalid usage instructions account for most unique true positives ($22/34$), often when documented workflows can be executed.
Among Codex Review's $44$ unique true positives, $14$ are typos or grammar mistakes and $9$ are formatting issues such as trailing whitespace.

The main exception to \approach{}'s otherwise strong per-repository results is React: Mini-SWE-agent finds more true positives ($7$ versus $3$), and \approach{} has low precision ($21\%$).
Manual inspection suggests that React favors execution-oriented checks: Mini-SWE-agent's true positives are mostly concrete execution failures, such as nonexistent \texttt{env.tryRecord()} and stale signatures.
\approach{} finds inconsistencies in ESLint-config, but also flags illustrative React documentation as non-exhaustive and benign terminology differences as contradictions.

\subsection{RQ2: Feedback from Real-World Developers}
\label{s:realworld}

To evaluate whether README bugs detected by \approach{} are actionable and useful to developers, we submitted alerts and patches from the main evaluation dataset and from an additional set produced by an earlier execution of \approach{} on the same six repositories.
This additional set contains $290$ README-changing commits, plus one non-README-changing commit selected by the commit filter.
We deduplicate alerts by location and group related alerts into one report, submitted either as an issue or as a PR (or, for Linux, as a patch submission).
To reduce maintainer burden, we keep at most two unconfirmed reports open per repository at any time; once maintainers confirm or fix a report, we may submit another one for that repository.

\begin{table}[!t]
    \caption{Reported, confirmed, and fixed README bugs.}
    \label{tab:results-inthewild}
    \centering
    \scriptsize
    \setlength{\tabcolsep}{2pt}
    \begin{tabular*}{\columnwidth}{@{}l@{\extracolsep{\fill}}rrr@{}}
        \toprule
        \textbf{Project} & \textbf{Reported} & \textbf{Confirmed} & \textbf{Fixed} \\
        \midrule
            Linux & 22 & 22 & 4 \\
            Spring Boot & 17 & 17 & 17 \\
            AutoGPT & 9 & 0 & 0 \\
            Ollama & 9 & 0 & 0 \\
            TensorFlow & 5 & 5 & 5 \\
            React & 4 & 0 & 0 \\
        \midrule
        \textbf{Total} & 66 & 44 & 26 \\
        \bottomrule
    \end{tabular*}
\end{table}

Table~\ref{tab:results-inthewild} shows that \approach{} finds actionable README bugs in the six evaluated projects.
We reported \readubugsizereport{} bugs\footnote{Refer to \url{https://doehyunbaek.github.io/readu-reports} for the complete list.}, received maintainer confirmation for \readubugsizeconfirm{} bugs, and \readubugsizefix{} of them have been fixed.
Overall, these responses indicate that \approach{} surfaces README bugs that maintainers consider actionable.
Reception was strongest in Spring Boot, Linux, and TensorFlow: Spring Boot maintainers confirmed and fixed all $17$ reports, Linux maintainers confirmed all $22$ reports and fixed $4$, and TensorFlow maintainers confirmed and fixed all $5$ reports.
The remaining $18$ confirmed but not-yet-fixed Linux reports have not been rejected; maintainer feedback indicates that they are delayed by normal upstream logistics, such as merge-window timing.

We have not yet received maintainer responses for AutoGPT, Ollama, and React.
In this small sample, the number of detected bugs does not clearly explain maintainer response: Table~\ref{tab:per-repo-effectiveness} shows many true positives for Ollama and AutoGPT ($120$ and $75$), but only a few for React ($3$).
This pattern may reflect project-specific review priorities, but the sample is too small for a definite conclusion.

\subsection{RQ3: Efficiency}
\label{s:efficiency}

The efficiency columns of Table~\ref{tab:big-evaluation} show that \approach{} remains practical despite its strong detection performance.
Compared with the two more expensive general-purpose baselines, Single LLM invocation and Codex Review, \approach{} is faster, uses fewer tokens, and is cheaper per commit, on average: $47.08$s versus $122.18$s--$130.93$s, $337$k versus $397$k--$504$k tokens, and \$0.0066 versus \$0.0073--\$0.0102.
Mini-SWE-agent is cheaper than \approach{} but this lower cost coincides with substantially lower true positives (63 versus 244) and precision (58\% versus 75\%).
This lower cost reflects a shallower exploration: Mini-SWE-agent averages $6.0$ LLM turns per commit, whereas \approach{}'s checker agents average $46.5$ turns combined once a commit passes the filter ($24.4$ internal and $22.2$ external).
Task-specific baselines are also cheaper, but they are much less effective.

\begin{figure}[t]
    \centering
    \includegraphics[width=\columnwidth]{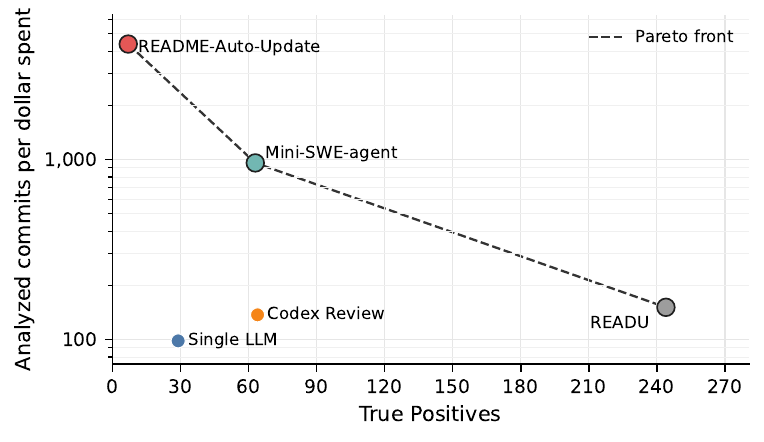}
    \caption{Pareto front of detection effectiveness and cost efficiency. The x-axis shows manually confirmed true positives; the y-axis shows commits analyzed per USD spent.}
    \label{fig:detection-pareto-front}
\end{figure}

Figure~\ref{fig:detection-pareto-front} places detection effectiveness and monetary efficiency on one Pareto plot.
The x-axis reports manually confirmed true positives, while the y-axis reports commits analyzed per dollar spent, the inverse of the average per-commit cost in Table~\ref{tab:big-evaluation}; points closer to the upper right are therefore better.
\approach{} is on the Pareto front because no baseline reaches its $244$ true positives at equal or higher cost efficiency.
Mini-SWE-agent and README-Auto-Update are cheaper per commit, but they trade that efficiency for much lower coverage.

\begin{figure}[t]
    \centering
    \includegraphics[width=\columnwidth]{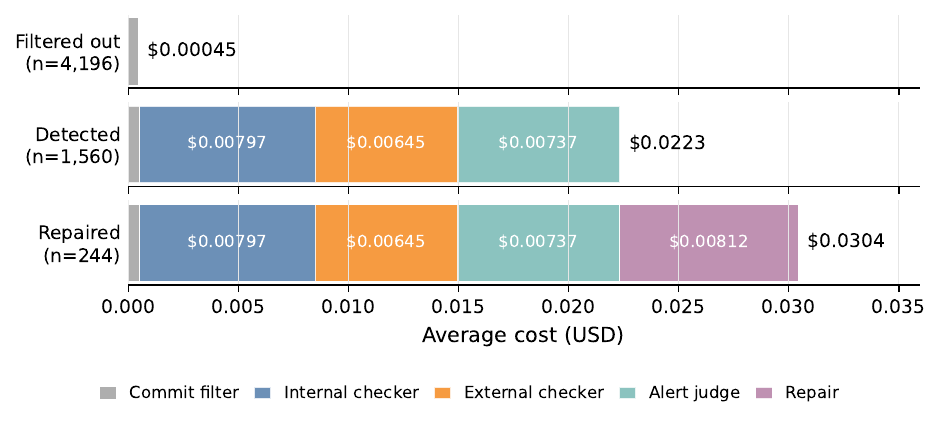}
    \caption{Average \approach{} cost by component for filtered-out commits, detected cases, and repaired targets.}
    \label{fig:phase-cost-breakdown}
\end{figure}

Figure~\ref{fig:phase-cost-breakdown} explains where this efficiency comes from.
For the $4{,}196$ commits discarded by the commit filter, \approach{} pays only the filter cost, averaging \$0.00045 per commit.
The remaining $1{,}804$ commits incur the detection stages: \$0.00052 for the filter, \$0.00797 for the internal checker, \$0.00645 for the external checker, and \$0.00737 for the alert judge, totaling \$0.0223 for each of the $1{,}560$ detection-only cases.
For the $244$ repaired targets, repair adds \$0.00812, bringing the end-to-end cost to \$0.0304 per target.
Thus, the commit filter keeps the overall per-commit cost low by preventing most commits from reaching the checker and judge stages.

\subsection{RQ4: Component Usefulness}
\label{s:ablation}

\subsubsection{Commit Filter}

\approach{} uses a commit filter to discard commits unlikely to introduce README bugs.
README-Auto-Update uses a similar component, called a relevance classifier, so we compare the two filters using two classifier-style outcomes: predicted positives (commits passed to the next stage) and false negatives (bug-inducing commits discarded by the filter).
Because we lack complete ground truth for the evaluated commits, we use the union of commits for which any evaluated approach raised a manually confirmed true-positive alert as a proxy for the set of commits with README bugs.

Table~\ref{tab:commit-filter} shows that the commit filter substantially reduces the search space while preserving almost all bug-inducing commits.
It marks $1{,}804$ of $6{,}000$ commits as predicted positives and yields only $9$ false negatives among the $268$ commits with README bugs.
In contrast, README-Auto-Update is much more aggressive, passing only $54$ commits but producing $255$ false negatives.
This highlights the tradeoff between precision and recall: we favor a high-recall commit filter because deciding from the commit diff alone whether a commit introduces a README bug is difficult, and the full pipeline's cost is acceptable.

\begin{table}[t]
    \caption{Commit-filter outcomes. Predicted positives are commits passed to downstream analysis; false negatives are proxy bug-inducing commits discarded by the filter.}
    \label{tab:commit-filter}
    \centering
    \footnotesize
    \resizebox{\columnwidth}{!}{
        \begin{tabular}{@{}l r@{\,/\,}l r@{\,/\,}l@{}}
            \toprule
            \textbf{Technique} & \multicolumn{2}{c}{\textbf{Predicted Positives}} & \multicolumn{2}{c}{\textbf{False Negatives}} \\
            \midrule
            README-Auto-Update & $54$ & $6{,}000$ & $255$ & $268$ \\
            \approach{} & $1{,}804$ & $6{,}000$ & $9$ & $268$ \\
            \bottomrule
        \end{tabular}
    }
\end{table}

\subsubsection{Internal Consistency Checker}

The ablation rows of Table~\ref{tab:big-evaluation} show that the internal consistency checker is essential for coverage.
Removing it reduces true positives from $\readutp{}$ to $74$ and precision from \readupr{} to 68\%, even though it roughly halves token use and cost.
Thus, internal repository checks provide most of the detection signal.

\subsubsection{External Consistency Checker}

The external consistency checker increases coverage at a modest precision cost.
Without it, \approach{} finds $204$ true positives with $80\%$ precision; adding it raises true positives to $\readutp{}$.
Thus, external consistency checker is useful supplement to the internal consistency checker, where it detects bug classes missable with internal checker alone.
Specifically, the external consistency checker alone raises $47$ of the $63$ true positives ($75\%$) in the incorrect external references category.

\subsubsection{Alert Judge}

The alert judge primarily improves precision.
Without it, \approach{} reports $275$ true positives but also $180$ false positives, yielding $60\%$ precision; with it, false positives drop to $81$ and precision rises to $75\%$.
Specifically, alert judge decreases false positives in the category of pre-existing issues from $125$ to $32$, showing that the alert judge is an effective strategy for removing these alerts which could be considered noisy to the maintainers.

\section{Threats to Validity}
\label{s:threat}

Our repository selection may bias the results toward popular and active projects, limiting generalizability to smaller or less active repositories.
Our commit selection may introduce temporal bias because it only considers commits from 2025-04-07 to 2026-04-07.
We focus on these repositories and commits because detecting README bugs in recent commits of popular repositories is potentially more impactful.
We run each baseline and our approach only once because of cost constraints; therefore, our results are subject to LLM nondeterminism.
We consider this acceptable because our evaluation covers a large number of commits (1,000) across multiple repositories (6), and the effectiveness metrics show substantial differences.
We use an open-weight model, DeepSeek V4 Flash, and do not evaluate current frontier proprietary models such as GPT-5.5 or Claude Opus 4.8, which could be more effective.
We exclude these models because of cost constraints and because efficiency is important for detecting and repairing README bugs, motivating the need for lighter, cheaper approaches.
Our monetary cost estimates are influenced by server-side caching behavior of the LLM inference providers.
Our evaluation results involving external resources may change as external environments evolve, e.g., because of link rot.

\section{Related Work}
\label{sec:related}

\paragraph{Empirical research on software documentation}

Prior empirical studies provide important context for understanding documentation quality and maintenance~\cite{DBLP:conf/doceng/ForwardL02,DBLP:conf/sigdoc/SouzaAO05,DBLP:journals/ese/PranaTTAL19,DBLP:conf/icse/AghajaniNVLMBL19,DBLP:journals/tse/GaoTZ25}.
We discuss two papers that are particularly relevant to this paper in detail.
Dagenais and Robillard~\cite{DBLP:conf/sigsoft/DagenaisR10} interviewed developers who write and read documentation, finding that keeping documentation synchronized with code changes can improve code quality.
This observation motivates \approach{} and similar automated techniques for detecting and repairing documentation bugs.
Aghajani et al.~\cite{DBLP:conf/icse/AghajaniNVLMBL19} mined open-source repositories, Stack Overflow, and developer mailing lists to derive a taxonomy of documentation issues.
Our work complements these empirical studies in two ways.
First, we conduct a comprehensive README bug-detection evaluation spanning 6,000 commits across six repositories and covering three general-purpose LLM-based techniques and two README-specific techniques.
This evaluation reveals the relative strengths and weaknesses of these techniques.
Second, by reporting detected bugs to maintainers and tracking confirmations and fixes, we provide evidence about the kinds of documentation bugs that occur in practice and how projects respond to them.
The results show substantial project-to-project variation in both bug prevalence and maintainer responsiveness.

\paragraph{Detection techniques for general documentation bugs}

Prior work has studied several forms of documentation bug detection.
Code-comment inconsistencies~\cite{DBLP:conf/scam/StulovaBGN20, DBLP:conf/acl/PanthaplackelNG20, DBLP:conf/aaai/PanthaplackelLG21, DBLP:conf/icse/RongYLTZSH25} and API-documentation errors~\cite{docref, DBLP:conf/icse/ZhouGCHPG17} have received particular attention; we discuss one representative example from each area.
Panthaplackel et al.~\cite{DBLP:conf/aaai/PanthaplackelLG21} train a neural just-in-time detector that flags when a code change makes its associated natural-language comment inconsistent.
Like that work, we use code changes as the trigger for just-in-time bug detection.
DOCREF~\cite{docref} detects API-documentation errors by combining NLP techniques with island parsing.
Specifically, DOCREF extracts code names from API-reference prose and examples, then reports mismatches when those names do not resolve to locally declared names or API elements in the latest generated reference.
These approaches target local consistency: comments are placed next to code, and DOCREF checks consistency within generated API-reference documentation.
In contrast, \approach{} targets repository-level documentation, where relevant documentation can appear anywhere in the repository and the facts needed to validate it may come from elsewhere in the repository or from external resources.
An interesting direction for future work is to study whether repository-level approaches like ours can also improve these more local documentation-bug settings.

\paragraph{Repository-level software engineering}

Recent work increasingly treats the repository, rather than an isolated file or snippet, as the unit of analysis for software-engineering automation.
One central task is repository-level code generation, where a model must use cross-file repository context to synthesize code consistent with the surrounding project~\cite{DBLP:conf/emnlp/ZhangCZKLZMLC23,DBLP:conf/icse/ZhengWSZM0Z25,DBLP:conf/icse/0001WGCZMZ25,liu2025reposcopeleveragingchainawaremultiview}.
Other repository-level work uses graph representation learning for security-patch detection~\cite{DBLP:conf/icse/WenL00W025} and LLM-based agents for Rust issue resolution~\cite{DBLP:journals/corr/abs-2602-22764}.
\approach{} shares this repository-level setup, which better matches realistic software-engineering tasks but requires gathering repository-wide evidence across code, configuration, and documentation effectively and efficiently.
The key difference is the target task: prior systems focus on code generation, patch detection, or issue resolution, whereas \approach{} uses repository-level agents specifically for just-in-time README bug detection and repair.

\paragraph{Detection techniques for README bugs}

DOCER~\cite{docer} and README-Auto-Update~\cite{readme-auto-update} are the closest prior works on README-bug detection.
DOCER targets README files and wiki pages: it uses regular expressions to extract code elements, then raises an alert when an extracted element's occurrences in the codebase drop from a positive number to zero, indicating a possible deletion or rename.
Due to its simple design, DOCER misses many of the diverse documentation bugs in our evaluation dataset.
README-Auto-Update is a five-step LLM pipeline for just-in-time README bug detection.
Although README-Auto-Update describes an ``agentic'' workflow, it remains a fixed pipeline for a narrower task: checking whether a target README file should be updated to stay consistent with a code change.
It is therefore less flexible than the three agentic approaches we evaluate, which can gather evidence and inspect repository context across files.
This design limits the kinds of README bugs it can detect in two ways.
First, README-Auto-Update focuses on a target README file rather than locating documentation bugs across repository-level documentation.
\approach{} handles documentation wherever it appears in the repository, and this broader scope accounts for most of the real-world bugs it catches.
Second, README-Auto-Update does not target README bugs caused by external factors such as dependencies.
Our external consistency checker specifically targets such cases.
Finally, our evaluation shows that \approach{} outperforms README-Auto-Update on the broader task of just-in-time detection of README bugs.

\section{Conclusion}
\label{sec:conclusion}

We present \approach, a technique for detecting and repairing README bugs.
Through the combination of high-recall commit filter, two checkers targeting internal and external consistency, and high-precision alert judge, \approach{} successfully detects many real-world documentation bugs.
In addition, \approach{} automatically synthesizes patches that resolves these bugs, alleviating the efforts of the maintainers.
We envision a future where techniques like \approach{} are integrated to software development life cycles of important projects, contributing to the usability and the maintainability of them.

\section*{Data Availability}
Our code and data are available at \url{https://github.com/sola-st/readu}.

\bibliographystyle{IEEEtran}
\bibliography{references}

\end{document}